\documentclass[prl,twocolumn,preprintnumbers]{revtex4}
\usepackage{amsmath}
\usepackage{graphicx}
\usepackage{subfigure}

\def\SphiK{S_{\phi K}}
\def\CphiK{C_{\phi K}}
\def\BtophiKs{B\to\phi K_S}
\def\Acp{A_{\rm CP}^{b\to s\gamma}}
\def\bsbsbar{B_s$--$\overline{B_s}}
\def\bbbar{B^0$--$\overline{B}^0}

\begin{document}
\preprint{KAIST-TH 2003/12}

\title{$\BtophiKs$ $CP$ asymmetries as a probe of supersymmetry\footnote{%
    Talk presented at ICFP2003, Seoul,
    Oct 6--11, 2003,
    to appear in the proceedings.
}}



% \author{G.L.~Kane}
% %\email[]{gkane@umich.edu}
% %\homepage[]{Your web page}
% %\thanks{}
% %\altaffiliation{}
% \affiliation{Michigan Center for Theoretical Physics, University of
% Michigan\\ Ann Arbor, MI 48109, USA}
% 
% \author{P.~Ko}
% %\email[]{pko@muon.kaist.ac.kr}
% %\homepage[]{Your web page}
% %\thanks{}
% \affiliation{Department of Physics, KAIST \\ Daejon 305-701, Korea}
% 
% \author{C.~Kolda}
% %\email[]{ckolda@nd.edu}
% %\homepage[]{Your web page}
% %\thanks{}
% %\altaffiliation{}
% \affiliation{Department of Physics, University of Notre Dame \\ 
% Notre Dame, IN 46556, USA}

\author{Jae-hyeon Park}
\email[]{jhpark@particle.kaist.ac.kr}
\affiliation{Department of Physics, KAIST \\ Daejeon 305-701, Korea}

% \author{Haibin Wang}
% %\email[]{haibinw@physics.purdue.edu}
% %\homepage[]{Your web page}
% %\thanks{}
% %\altaffiliation{}
% \affiliation{Department of Physics, Purdue University \\
% West Lafayette, IN 47907, USA}
% 
% \author{Lian-Tao Wang}
% %\email[]{liantaow@pheno.physics.wisc.edu}
% %\homepage[]{Your web page}
% %\thanks{}
% %\altaffiliation{}
% \affiliation{Department of Physics, University of Wisconsin  \\
% Madison, WI 53706, USA}


%\date{\today}

\begin{abstract}
We consider sparticle contributions to time dependent $CP$ asymmetry
in $\BtophiKs$.
As for the gluino-squark loop, $LR$ or $RL$ insertion
is more likely to give recently observed $\SphiK < 0$ than 
$LL$ or $RR$ insertion.
Neutral Higgs contribution does not change $\SphiK$ very much
due to the strong constraint from $B_s \rightarrow \mu^+ \mu^-$.
We also show correlations among related processes
such as direct $CP$ asymmetry in $B \rightarrow X_s \gamma$ 
and $\bsbsbar$ mixing,
and discuss theoretical motivations for desired 
strength of flavor violation.
\end{abstract}

% \pacs{12.60.Jv, 11.30.Er}
%\keywords{}

\maketitle

% Introduction

%% SM prediction, data, difference
%% Sensitivity to NP
%% contributions considered
%\noindent 
Time dependent $CP$ asymmetry in $\BtophiKs$ mode
is written as
\begin{align*}
  {\cal A}_{\phi K} (t) &\equiv 
  {{\Gamma (\overline{B}^0 (t) \rightarrow \phi K_S ) - 
      \Gamma (      B^0 (t) \rightarrow \phi K_S )   } \over
    {\Gamma (\overline{B}^0 (t) \rightarrow \phi K_S ) + 
      \Gamma (      B^0 (t) \rightarrow \phi K_S )   }} \\
  &= - C_{\phi K} \cos ( \Delta m_d t )
  + S_{\phi K} \sin ( \Delta m_d t ) ,
\end{align*}
where $\Delta m_d$ denotes the mass splitting between
the two mass eigenstates composed of $B^0$ and $\overline{B}^0$.
% The coefficients of the sine and cosine terms are written as
% \begin{equation}
%   C_{\phi K}  = 
%     { 1 - | \lambda_{\phi K} |^2 \over 1 + | \lambda_{\phi K} |^2 } ,\quad
%     S_{\phi K}  =  
%     { 2~ {\rm Im} \lambda_{\phi K} \over 1 + | \lambda_{\phi K} |^2 } ,
% \end{equation}
% in terms of the parameter $\lambda_{\phi K}$ defined by
% \begin{equation}
%   \lambda_{\phi K} \equiv - e^{ - 2 i (\beta + \theta_d )} 
%   {\overline{A} (\overline{B}^0 \rightarrow \phi K_S ) \over 
%     A (B^0 \rightarrow \phi K_S ) } .
% \end{equation}
Current data on the coefficients of the sine and cosine terms
and their Standard Model (SM) predictions
are summarized in Table~\ref{tab:data}.
\begin{table}
  \centering
\begin{tabular}{c|c|c}
  \hline \hline
  & $S_{\phi K}$ & $C_{\phi K}$ \\ \hline
  BaBar\cite{browder} & $+0.45 \pm 0.43 \pm 0.07$ & $-0.38 \pm 0.37 \pm 0.12$ \\
  Belle\cite{Abe:2003yt} & $-0.96 \pm 0.50^{+0.09}_{-0.11}$ & $+0.15 \pm 0.29 \pm 0.07$ \\
  Average & $-0.15 \pm 0.33$ & $-0.05 \pm 0.24$ \\
  SM & $0.734 \pm 0.054$ & 0 \\
  $\text{Avg.} - \text{SM}$ & $- 2.7 \sigma$ & $- 0.2 \sigma$ \\
  \hline \hline
\end{tabular}
  \caption{Current status of time dependent $CP$ asymmetry in $\BtophiKs$.}
  \label{tab:data}
\end{table}
It shows a 2.7 $\sigma$ discrepancy between the measured value and
SM prediction of $\SphiK$.
This has become a hot issue in the high energy physics community recently.
One of the reasons why physicists are focusing on $\BtophiKs$ mode
is that it is loop suppressed in the SM because it is a purely 
flavor changing neutral current (FCNC) process.
This means that it is more sensitive to possible new physics effects
which presumably arise at loop level, than those processes
that have tree level contribution such as $B \rightarrow J/\psi K_S$.
In this work, we consider a possibility of filling this gap
between the data and SM value,
with effects from supersymmetry (SUSY).

\begin{figure}
  \centering
  \includegraphics{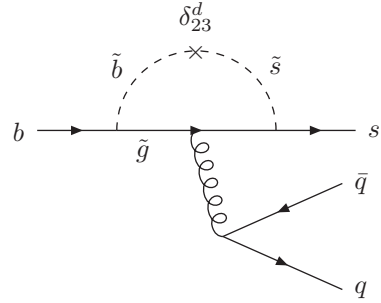}
  \caption{Gluino-squark loop penguin diagram.}
  \label{fig:squarkloop}
\end{figure}

% Analysis
%% MIA
%% constraints
%% explanation on RL dominance
%% Photon polarization
One major source of flavor violation in the general minimal
supersymmetric standard model (MSSM) is
non-diagonal squark mass matrix.
An off-diagonal element in the squark mass matrix in the super CKM
basis can constitute a diagram leading to $b \rightarrow s \bar{q} q$
through gluino mediation,
where $q = u,d,s,c,b$.
An example of this type of 
QCD penguin diagram is shown in Fig.~\ref{fig:squarkloop}.
\begin{figure*}[htbp]
  \centering
  \subfigure[Allowed region for $( \delta_{LL}^d )_{23}$]
  {\includegraphics[height=5cm]{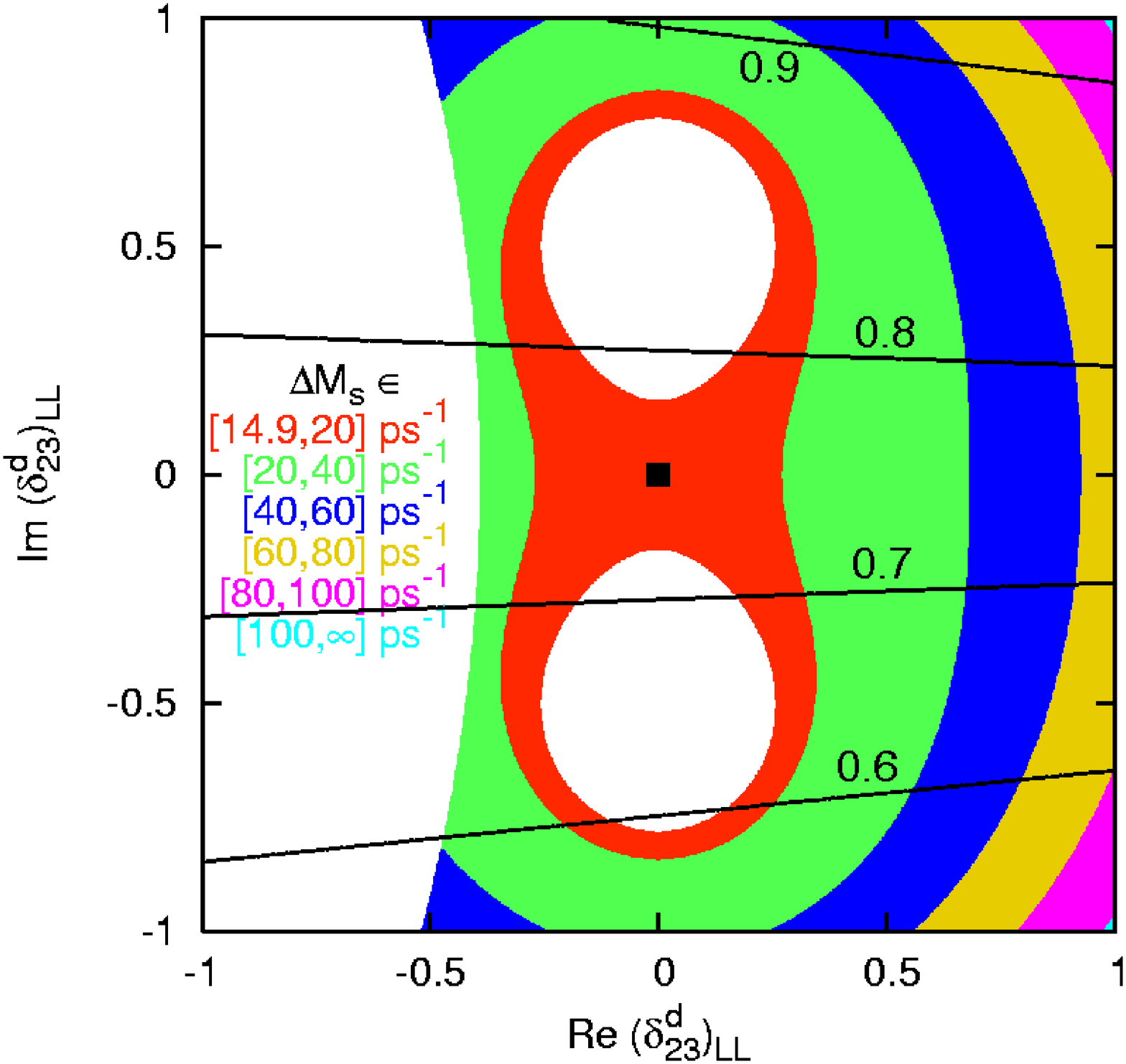}}
  \subfigure[$\SphiK$ vs.\ $\CphiK$]
  {\includegraphics[height=5cm]{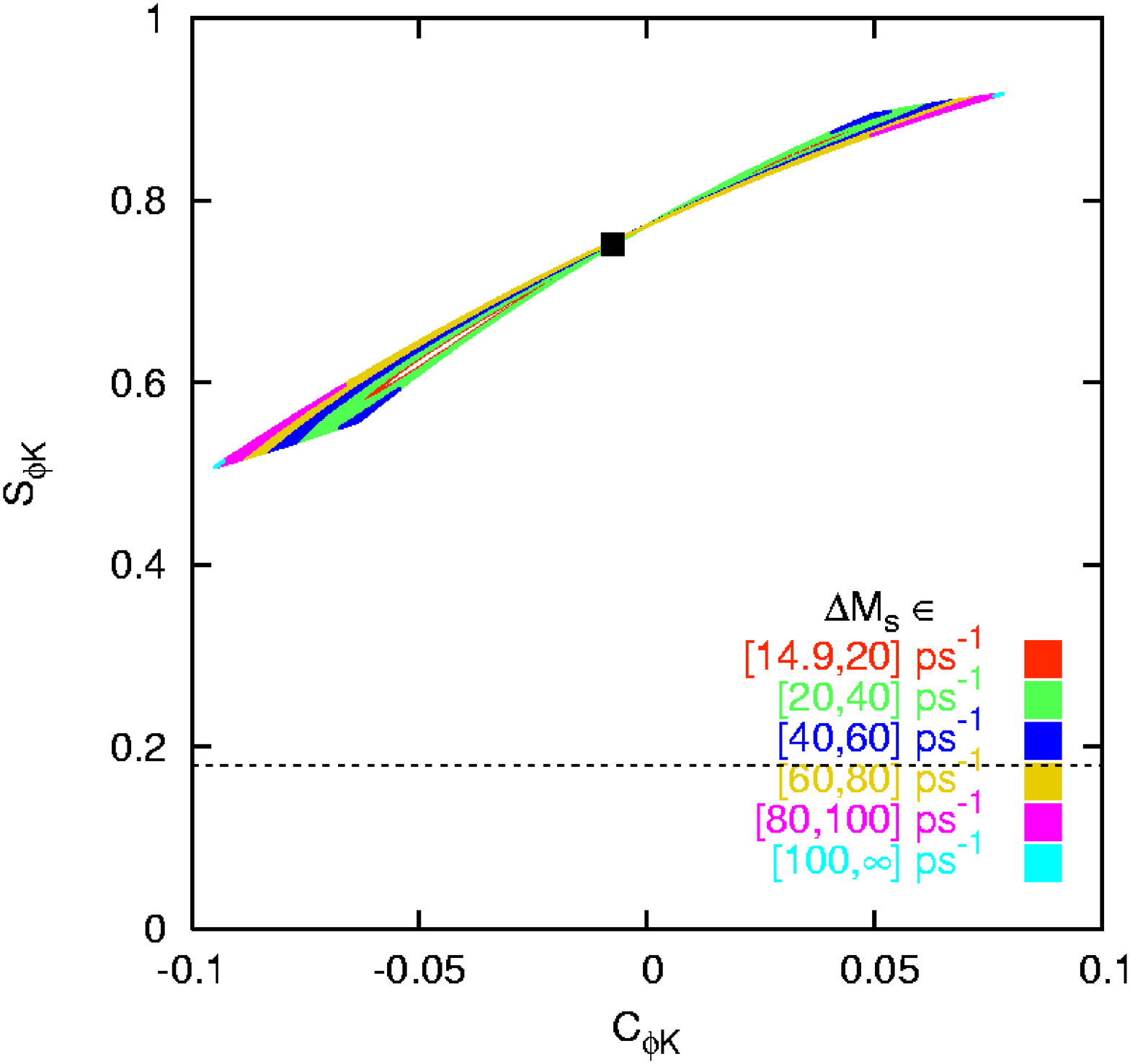}} 
  \\
  \subfigure[$\SphiK$ vs.\ $\Delta M_s$]
  {\includegraphics[height=5cm]{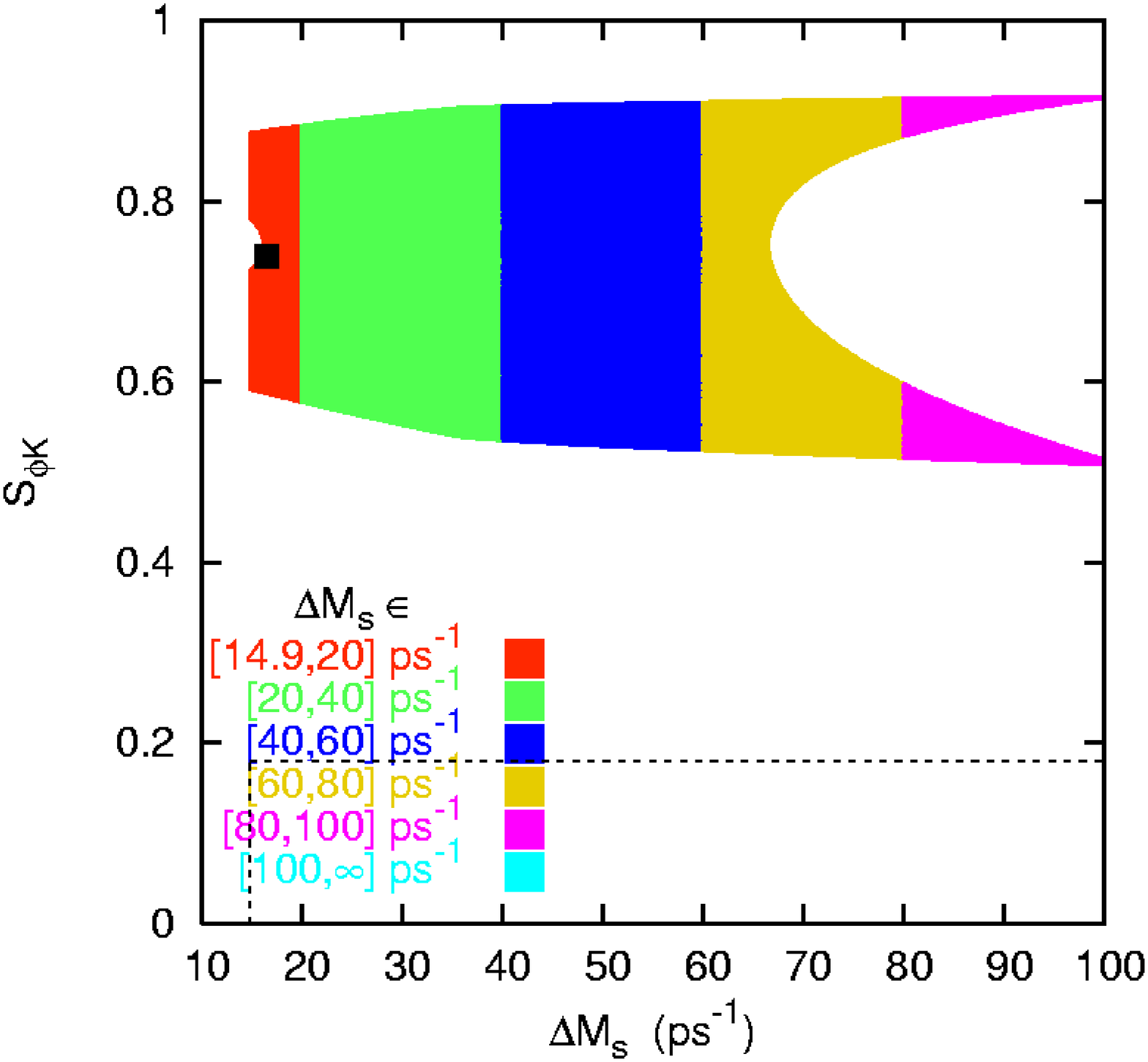}}
  \subfigure[$\SphiK$ vs.\ $\sin 2 \beta_s$]
  {\includegraphics[height=5cm]{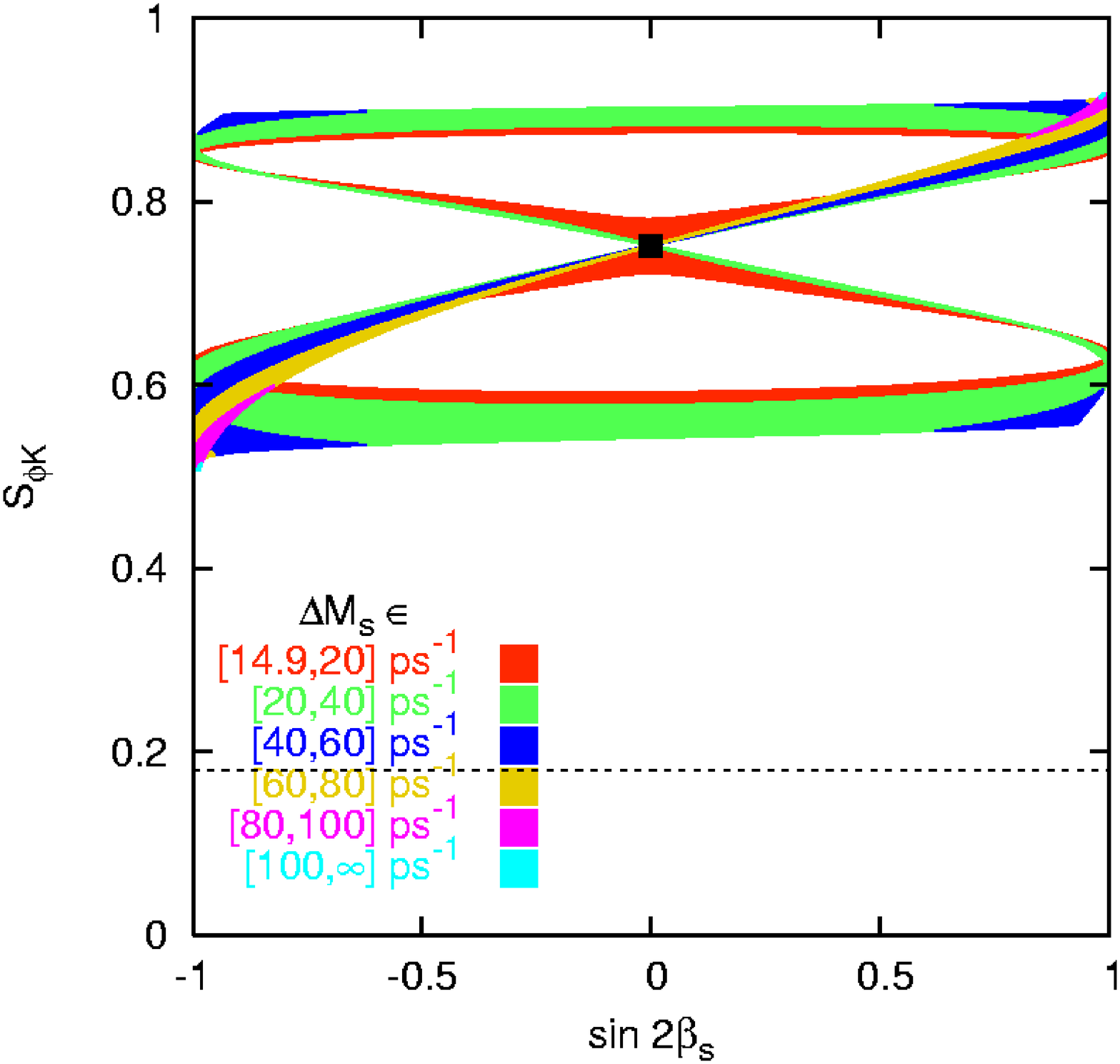}}
  \caption{(a) Allowed region on the complex plane
    of $( \delta_{LL}^d )_{23}$, and (b--d)
    correlations among observables for the $LL$ insertion case.
    Different shades are used for different values of $\Delta M_s$.
    The black square corresponds to the SM case.
    The dotted line represents the upper bound on 
    $\SphiK$ at 1-$\sigma$ level.
    Lines in Fig.~(a) are contours of $\SphiK$.}
  \label{fig:LLDB2}
\end{figure*}
Here we adopt the mass insertion (MI) approximation which is
one of the most convenient ways to analyzing this type of contribution.
We consider four MI's relevant to $b \rightarrow s$ transitions, i.e.,
$( \delta_{LL}^d )_{23}$, $( \delta_{RR}^d )_{23}$,
$( \delta_{LR}^d )_{23}$, and $( \delta_{RL}^d )_{23}$, one at a time.
In addition to these four cases,
we include another one called $RL$ dominance scenario \cite{Everett:2001yy}.
In this case,
$C_{7 \gamma}$ and $C_{8 g}$ are assumed to vanish
at the $m_b$ scale, so the observed decay of $B\rightarrow X_s \gamma$
must come from $\widetilde{C}_{7 \gamma}$, the chirality-flipped
version of $C_{7 \gamma}$.
This unconventional scenario and the usual case can be distinguished
by measuring the photon polarization.
Therefore $( \delta_{RL}^d )_{23}$ is forced to be a finite size
to account for the data unlike the other four cases,
and this corresponds to an extreme case 
with maximized SUSY effect in some sense.
In the following,
we turn on one of the four MI parameters at a time,
and scan over it imposing constraints from
$B \rightarrow X_s \gamma$ and $\bsbsbar$ mixing as follows:
\begin{align*}
  2.0 \times 10^{-4} < B ( B\rightarrow X_s \gamma ) < & 4.5 \times 10^{-4}, \\
  \Delta M_s > 14.9 \ \text{ps}^{-1}& \mbox{\cite{Stocchi:2000ps}}.
\end{align*}
We impose a rather generous bound on $B ( B\rightarrow X_s \gamma )$
to take into account theoretical uncertainties.
The common squark mass $\tilde{m}$ and the gluino mass $m_{\tilde{g}}$
are chosen to be $m_{\tilde{g}} = \tilde{m} = 400$ GeV.
We use QCD factorization \cite{Beneke:1999br}
in evaluating hadronic matrix elements
for $\BtophiKs$.
We do not assume any new physics in the $\bbbar$ mixing because
$\sin 2 \beta$ measurement from the $B \rightarrow J/\psi K_S$ mode
is well consistent with the SM fit.

\begin{figure*}[htbp]
  \centering
  \subfigure[Allowed region for $( \delta_{LR}^d )_{23}$]
  {\includegraphics[height=5cm]{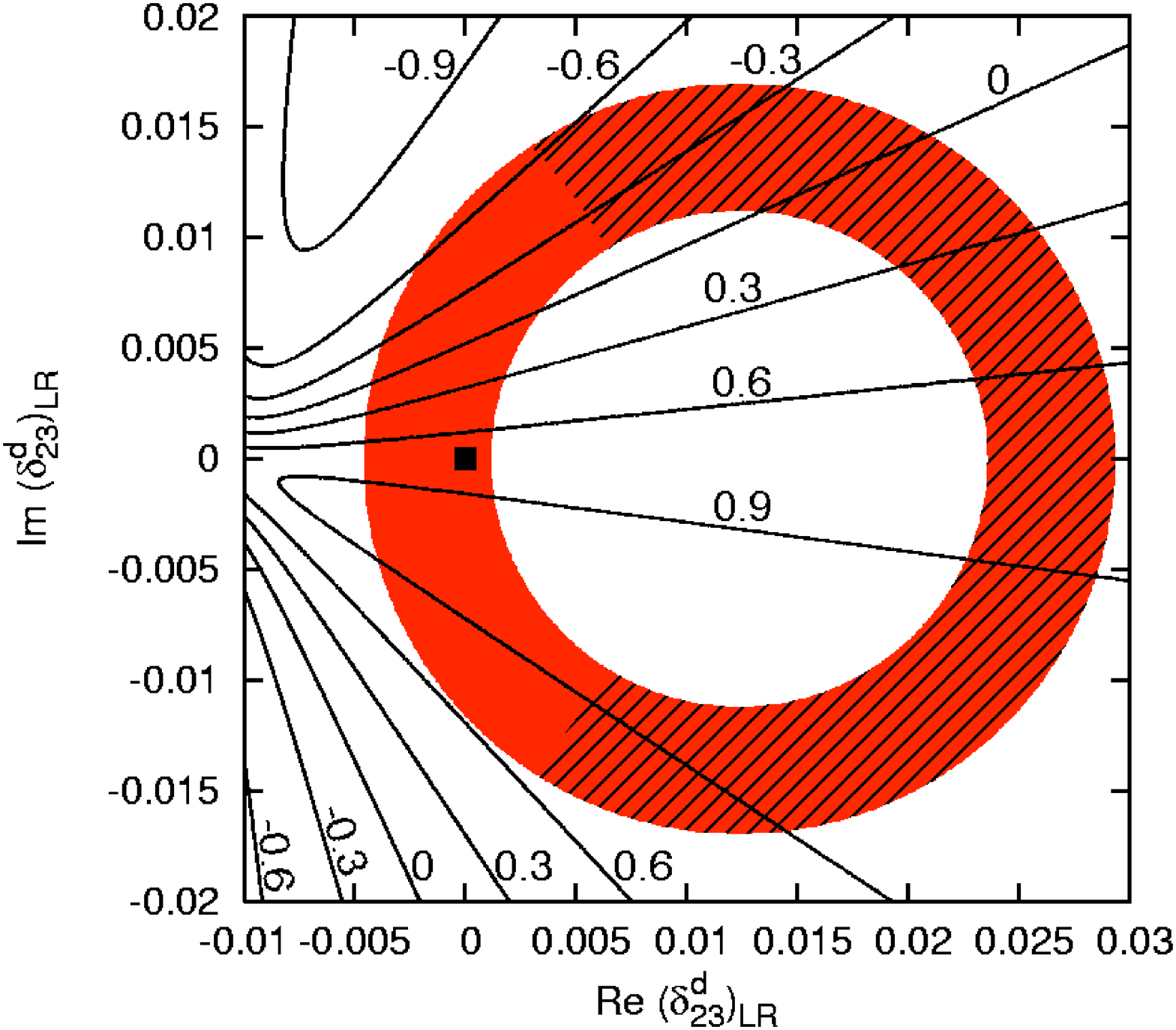}}
  \subfigure[$\SphiK$ vs.\ $\CphiK$]
  {\includegraphics[height=5cm]{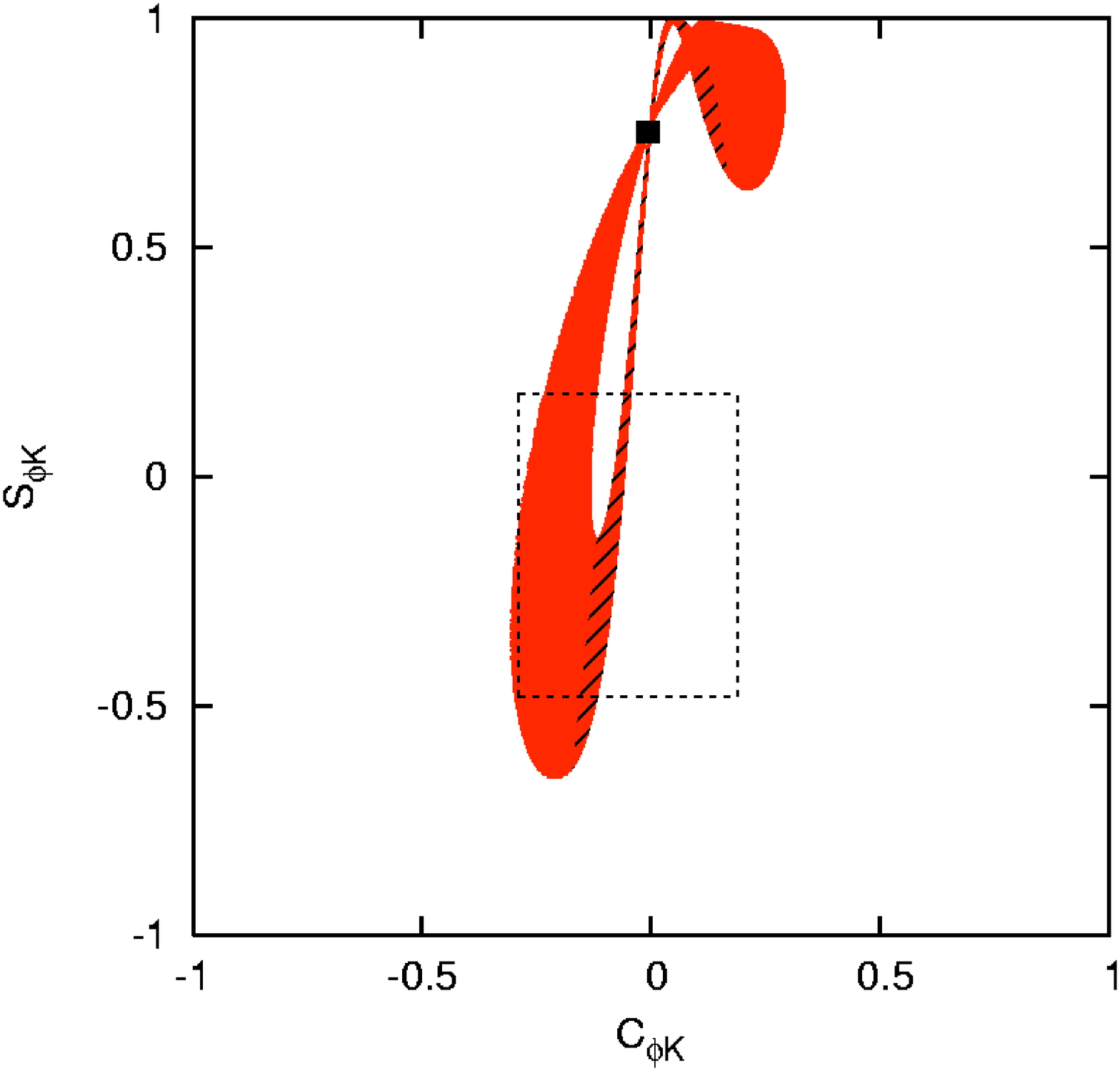}}
  \\
  \subfigure[$\SphiK$ vs.\ $B ( \BtophiKs )$]
  {\includegraphics[height=5cm]{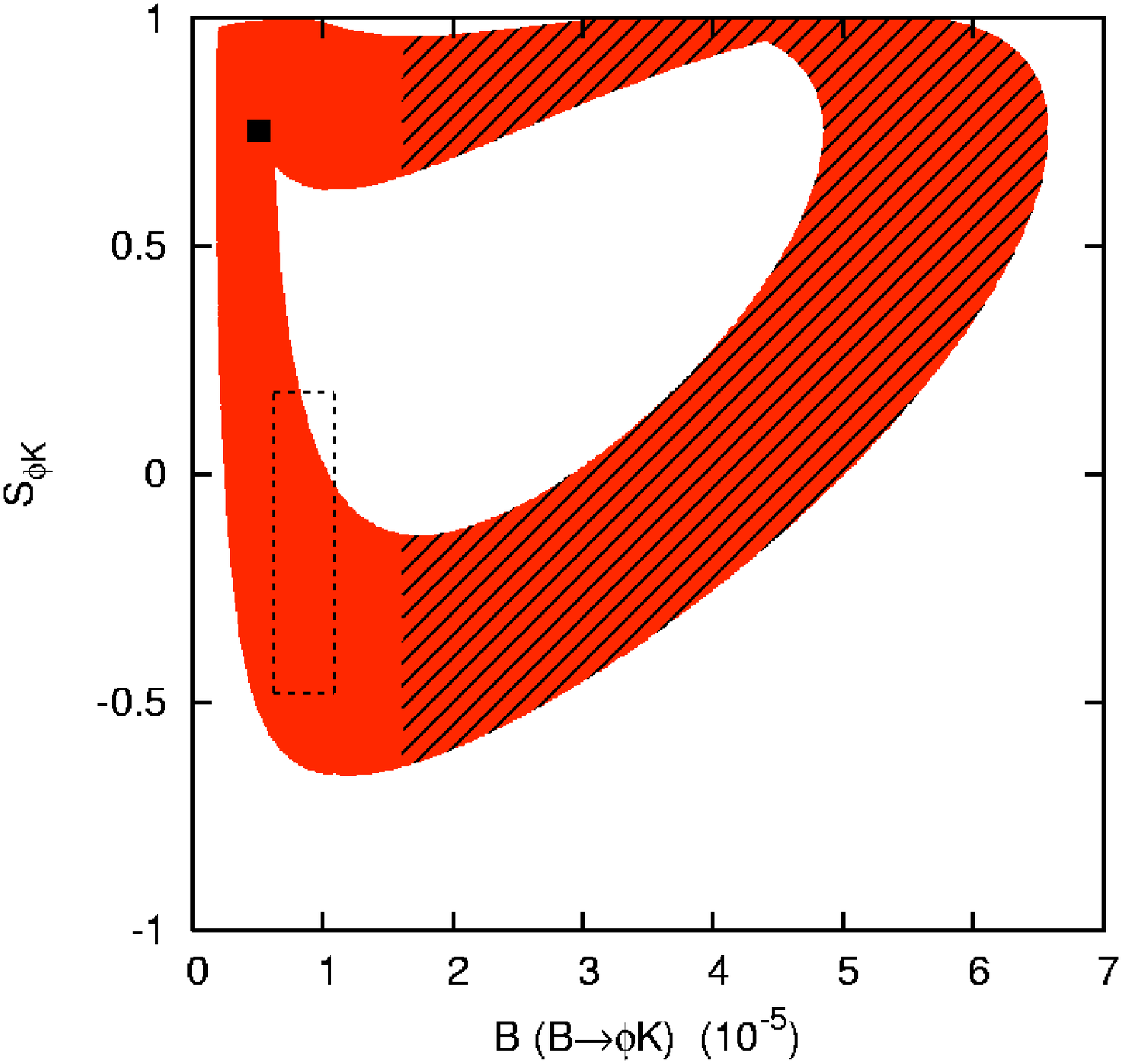}}
  \subfigure[$\SphiK$ vs.\ $\Acp$]
  {\includegraphics[height=5cm]{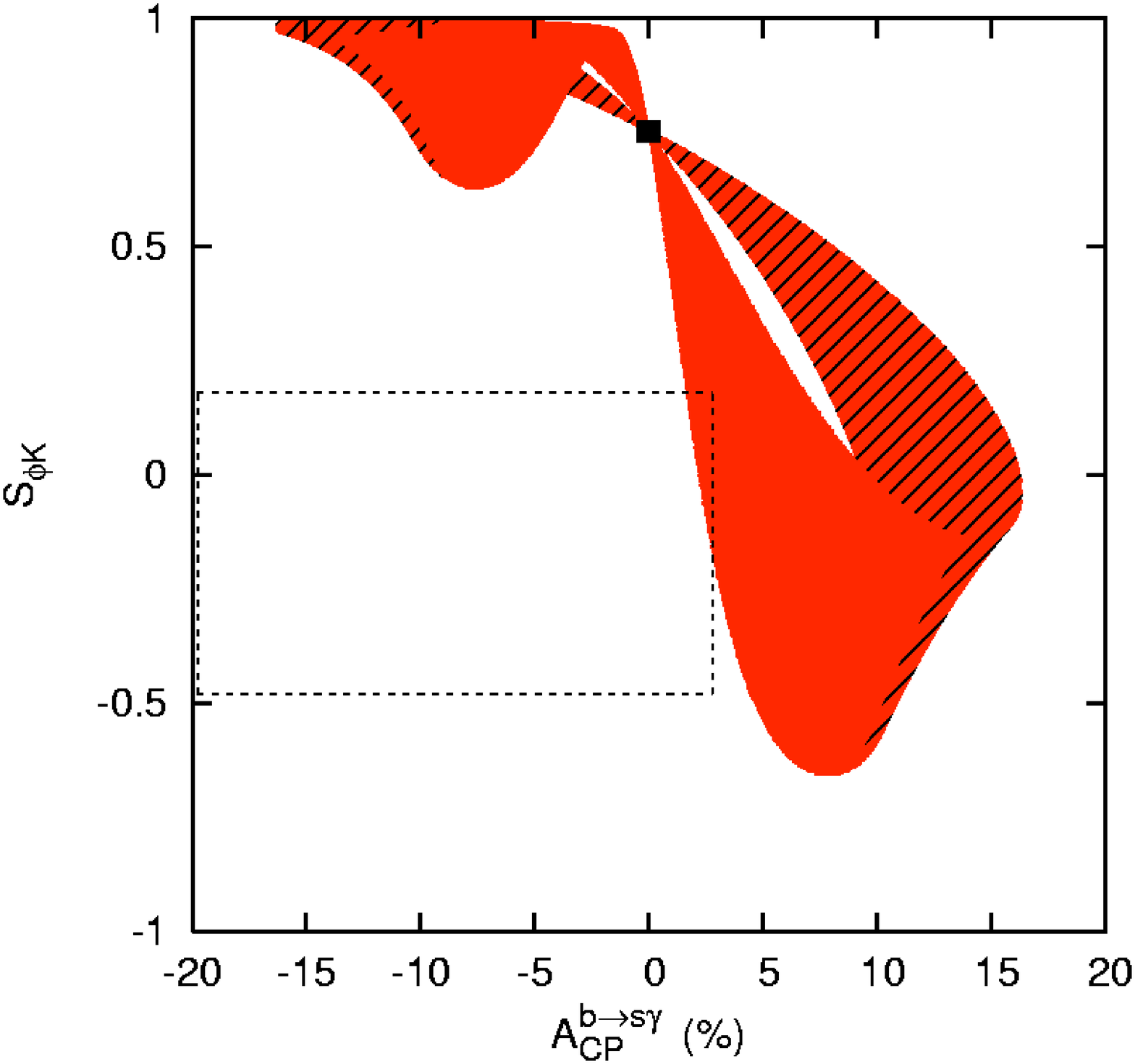}}
  \caption{(a) Allowed region on the complex plane
    of $( \delta_{LR}^d )_{23}$, and (b--d)
    correlations among observables for the $LR$ insertion case.
    The black square corresponds to the SM case.
    The dotted boxes represent current measurements at 1-$\sigma$ level.
    Lines in Fig.~(a) are contours of $\SphiK$.}
  \label{fig:LRDB1}
\end{figure*}
First, let us consider the $LL$ insertion case.
% We show plots for the $LL$ insertion case in
% Figs.~\ref{fig:LLDB2}.
Fig.~\ref{fig:LLDB2}~(a) is the region of $( \delta_{LL}^d )_{23}$
consistent with 
$B (B \rightarrow X_s \gamma)$ and
$\Delta M_s$ constraints.
The solid lines are contours of $\SphiK$.
The small square at the origin corresponds to the SM
case, because there is no SUSY contribution there.
The white space on the left is excluded by too small value of 
$B (B \rightarrow X_s \gamma)$.
The two holes around the $\mathrm{Re} ( \delta_{LL}^d )_{23} = 0$ line
are excluded by too small $\Delta M_s$.
In Fig.~\ref{fig:LLDB2}~(b), we plot
predictions of $\SphiK$ and $\CphiK$ from the allowed values of
$( \delta_{LL}^d )_{23}$.
The dotted horizontal line is the current 1-$\sigma$
upper limit on $\SphiK$.
This figure shows that the minimal value of $\SphiK$ 
in this case is about +0.5, and
the change of $\SphiK$ from the SM prediction is not significant.
% mention other observables not shown here
Although $\SphiK$ is not very much affected,
large effect in $\bsbsbar$ can be expected.
For example, we show the prediction of $\Delta M_s$ in 
Fig.~\ref{fig:LLDB2}~(c).
The SM value of $\Delta M_s$ is about 16 $\mathrm{ps}^{-1}$,
marked by the black square.
We can read huge enhancement of $\Delta M_s \sim 50\ \mathrm{ps}^{-1}$
is possible.
% connection with experiment
Also, we show $\sin 2 \beta_s$
in Fig.~\ref{fig:LLDB2}~(d), where $2 \beta_s$ is the phase of 
$\langle B_s | H_\mathrm{eff}^{\Delta B = 2} | \overline{B_s} \rangle$.
The SM prediction of it is almost zero, but
it can have any value between $-1$ and 1 here, which means
that $CP$ violation in $\bsbsbar$ mixing can drastically change.

% Comment on RR insertion
The $RR$ insertion case is almost identical to
the $LL$ insertion case.
The only difference is the way $B (B \rightarrow X_s \gamma)$ constrains
$( \delta_{RR}^d )_{23}$.
The $LL\ (RR)$ insertion mainly contributes
to $C_{7 \gamma}\ (\widetilde{C}_{7 \gamma})$, and
the dependence is such that
\begin{align*}
B (B \rightarrow X_s \gamma)
&\propto
|C_{7 \gamma}^\mathrm{SM} + C_{7 \gamma}^\mathrm{SUSY} |^2,
\quad
\text{for\ } LL,
\\
% \intertext{in the $LL$ insertion case, and}
B (B \rightarrow X_s \gamma) 
&\propto
|C_{7 \gamma}^\text{SM}|^2 + 
      |\widetilde{C}_{7 \gamma}^\text{SUSY}|^2, \quad 
\text{for\ } RR.
\end{align*}
%in the $RR$ insertion case.
Because of this difference, $( \delta_{RR}^d )_{23}$ plane does not
show a region excluded by $B (B \rightarrow X_s \gamma)$
that is present in Fig.~\ref{fig:LLDB2}~(a).
However, this additional parameter space does not help
shifting down $\SphiK$ very much.
Plots for this case are available in Ref.~\cite{Kane:2002sp}.

% \begin{figure*}[htbp]
%   \centering
%   \subfigure[Allowed region for $( \delta_{RR}^d )_{23}$]
%   {\includegraphics[height=6cm]{deltaRR.eps}}
%   \subfigure[$\SphiK$ vs.\ $\CphiK$]
%   {\includegraphics[height=6cm]{SC-RR.eps}}
%   \caption{(a) Allowed region on the complex plane
%     of $( \delta_{RR}^d )_{23}$, and (b)
%     correlation between $\SphiK$ and $\CphiK$ for the $RR$ insertion case.
%     Different shades are used for different values of $\Delta M_s$.
%     The black square corresponds to the Standard Model only case.
%     The dotted line represents the upper bound on 
%     $\SphiK$ at 1-$\sigma$ level.
%     Lines in Fig.~(a) are contours of $\SphiK$.}
%   \label{fig:RRDB1}
% \end{figure*}

It is well known that the chirality-flipping
$LR$ and $RL$ insertions receive
enhancement by the factor of $m_{\tilde{g}} / m_b$,
when they contribute to a (chromo) magnetic dipole operator.
This is why they are strongly constrained by $B \rightarrow X_s \gamma$.
For the same reason, they are more effective in modifying
$\SphiK$ and/or $\CphiK$ than $LL$ or $RR$ insertion
given the same magnitude of the MI parameter.
% Unlike the $LL$ or $RR$ insertion,
% $LR$ insertion can give large shift in $\SphiK$.
% The results are displayed in Figs.~\ref{fig:LRDB1}.
We show the region of $( \delta_{LR}^d )_{23}$ consistent with
$B (B \rightarrow X_s \gamma)$ and
$\Delta M_s$ in Fig.~\ref{fig:LRDB1}~(a).
As in the previous case,
the solid lines are contours of $\SphiK$, and
the small square at the origin corresponds to the SM
case.
The annulus with radius $\sim 10^{-2}$ comes from
the $B (B \rightarrow X_s \gamma)$.
Here the $B (B \rightarrow X_s \gamma)$ constraint is so strong that
$\Delta M_s$ does not change very much from the
SM value and play no role in constraining
$( \delta_{LR}^d )_{23}$.
We show $\SphiK$ and $\CphiK$ from these values of $( \delta_{LR}^d )_{23}$
in Fig.~\ref{fig:LRDB1}~(b).
The dotted box is 1-$\sigma$ intervals of these observables,
and we find some region where both $\SphiK$ and $\CphiK$
are within the box.
This plot also shows a definite correlation between them.
For $\SphiK$ less (bigger) than the SM value, $\CphiK <(>)\ 0$.
Fig.~\ref{fig:LRDB1}~(c) is the plot of $B(\BtophiKs)$,
with hatches on the region where $B(\BtophiKs) > 1.6 \times 10^{-5}$.
This region, excluded by $B(\BtophiKs)$ constraint, corresponds
to the hatched region in Fig.~\ref{fig:LRDB1}~(a).
Here we learn that significant portion of the $( \delta_{LR}^d )_{23}$
parameter space that is consistent with
$B (B \rightarrow X_s \gamma)$,
is excluded by $B(\BtophiKs)$.
In Fig.~\ref{fig:LRDB1}~(d) is displayed the correlation between
$\SphiK$ and direct $CP$ asymmetry in $B \rightarrow X_s \gamma$.
For $\SphiK$ less (bigger) than the SM value, $\Acp >(<)\ 0$.
It appears that $\Acp$ is outside the dotted box,
but the experimental uncertainty is still large, so we
cannot definitely conclude that this scenario is disfavored currently.

% \begin{figure}[htbp]
%   \centering
%   \subfigure[Allowed region for $( \delta_{RL}^d )_{23}$]
%   {\includegraphics[height=6cm]{deltaRL2.eps}}
%   \subfigure[$\SphiK$ vs.\ $\CphiK$]
%   {\includegraphics[height=6cm]{SC-RL2.eps}}
%   \subfigure[$\SphiK$ vs.\ $B ( \BtophiKs )$]
%   {\includegraphics[height=6cm]{SB-RL2.eps}}
%   \caption{(a) Allowed region on the complex plane
%     of $( \delta_{RL}^d )_{23}$, and (b--c)
%     correlations among observables for the $RL$ insertion case.
%     The black square corresponds to the Standard Model only case.
%     The dotted boxes represent current measurements at 1-$\sigma$ level.
%     Lines in Fig.~(a) are contours of $\SphiK$.}
%   \label{fig:RLDB1}
% \end{figure}

\begin{figure*}[htbp]
  \centering
  \subfigure[Allowed region for $( \delta_{RL}^d )_{23}$]
  {\includegraphics[height=5cm]{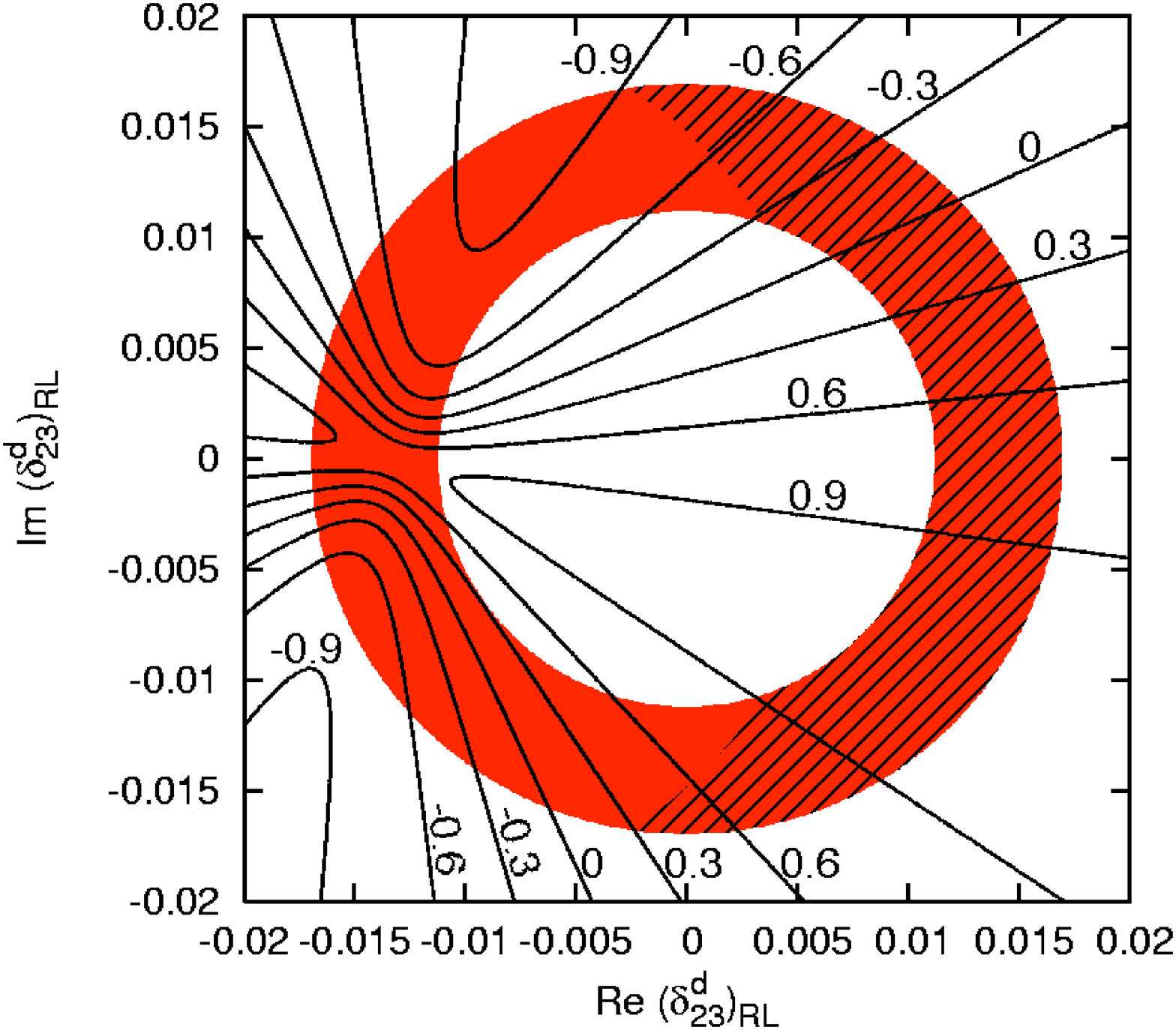}}
  \subfigure[$\SphiK$ vs.\ $\CphiK$]
  {\includegraphics[height=5cm]{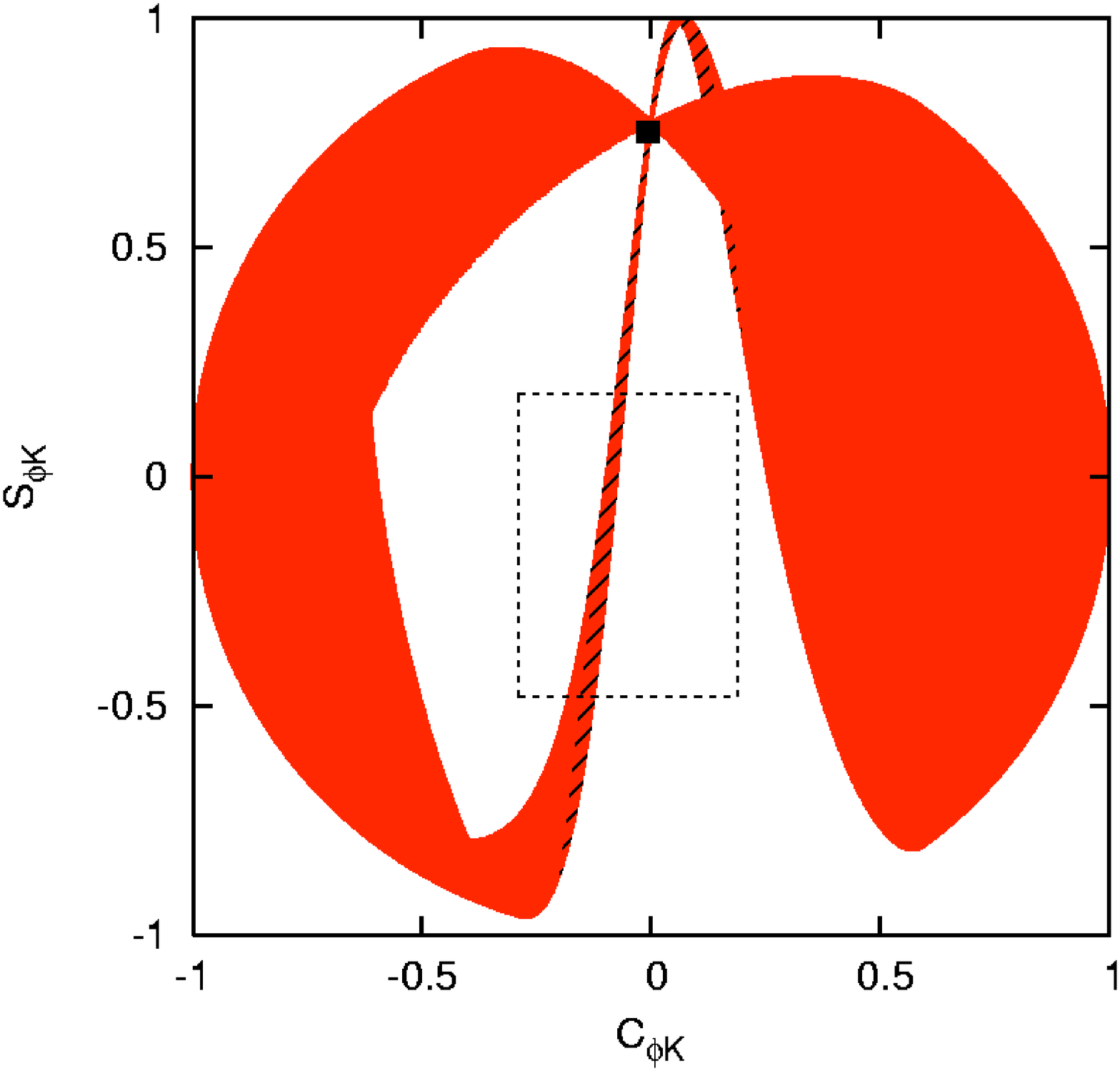}}
  \subfigure[$\SphiK$ vs.\ $B ( \BtophiKs )$]
  {\includegraphics[height=5cm]{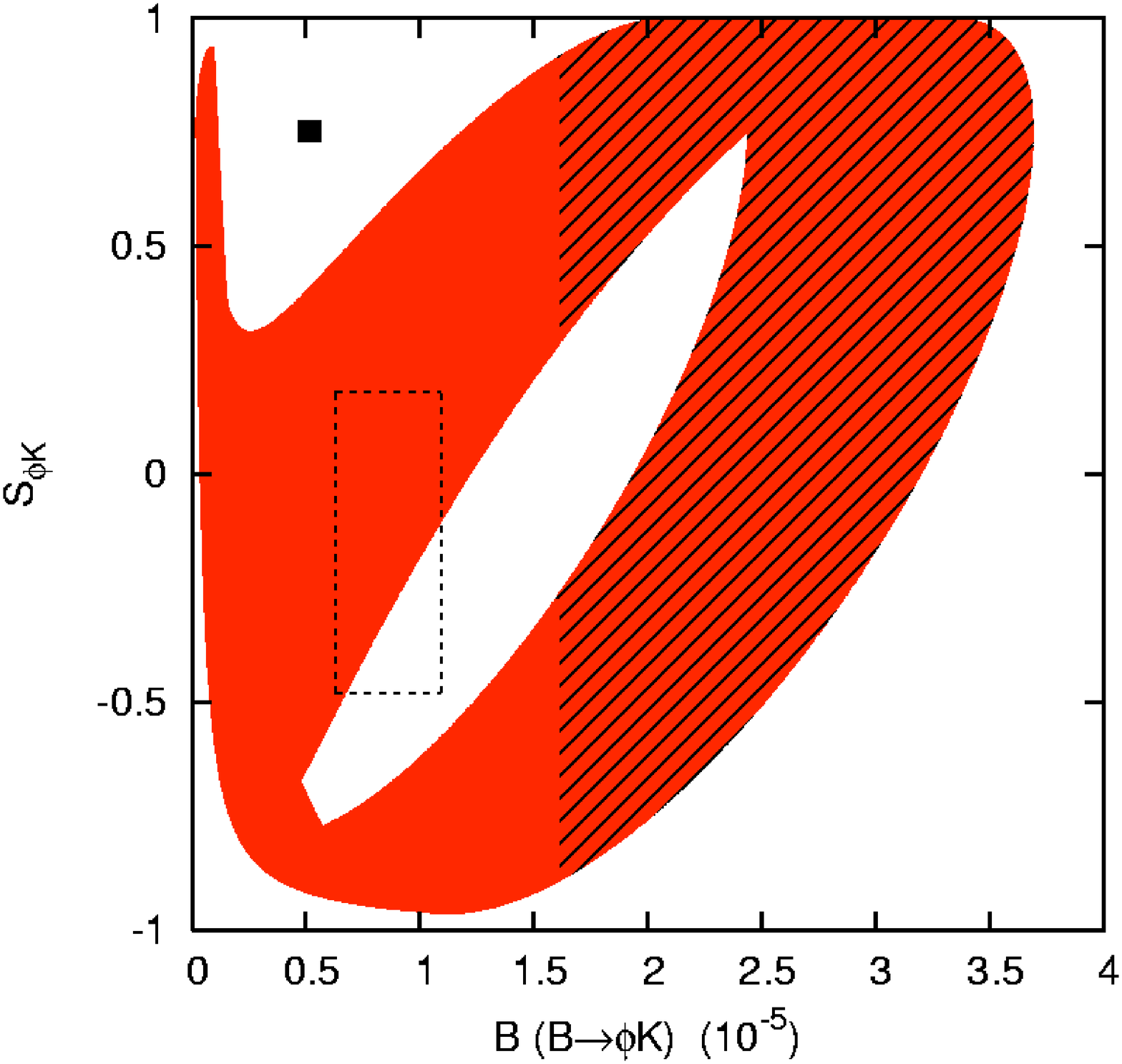}}
  \caption{(a) Allowed region on the complex plane
    of $( \delta_{RL}^d )_{23}$, and (b--c)
    correlations among observables for the $RL$ dominance case.
    The black square corresponds to the SM case.
    The dotted boxes represent current measurements at 1-$\sigma$ level.
    Lines in Fig.~(a) are contours of $\SphiK$.}
  \label{fig:RLdDB1}
\end{figure*}
We turn to the $RL$ dominance scenario.
%, with results shown in 
%Figs.~\ref{fig:RLdDB1}.
Fig.~\ref{fig:RLdDB1}~(a) is the region of $( \delta_{RL}^d )_{23}$ 
constrained by $B (B \rightarrow X_s \gamma)$ and
$\Delta M_s$.
Again, the solid lines are contours of $\SphiK$.
In this case, we are assuming that SM contributions to
$C_{7 \gamma}$ and $C_{8 g}$ are somehow canceled by
SUSY contributions.
Because we are always involving
SUSY contributions in $C_{7 \gamma}$ and $C_{8 g}$,
there cannot be any point on the $( \delta_{RL}^d )_{23}$ plane
that reduces to the SM case.
The allowed annulus, centered at the origin,
has radius $\sim 10^{-2}$, which is fixed by 
$B (B \rightarrow X_s \gamma)$.
As in the previous case,
the region consistent with $B (B \rightarrow X_s \gamma)$
is always consistent with $\Delta M_s$ as well.
Predictions of $\SphiK$ and $\CphiK$ from the allowed region
are shown in Fig.~\ref{fig:RLdDB1}~(b).
We can fit $\SphiK$ to any value between $-1$ and 1.
Moreover, big change in $\CphiK$ from the SM value of zero,
is expected in general, resulting
in any value between $-1$ and 1.
However, some region around the center is not covered.
In Fig.~\ref{fig:RLdDB1}~(c), we plot $B(\BtophiKs)$,
with hatches on the region with excessive $B(\BtophiKs)$.
Since there is only one weak phase in the
$b \rightarrow s \gamma$
amplitude, we have $\Acp = 0$ in this case.
If we turned on another weak phase such as coming from
$( \delta_{RR}^d )_{23}$, we could have non-vanishing $\Acp$,
but we do not pursue more detailed analysis here.

% Comment on usual RL
We also analyzed the usual $RL$ insertion case,
keeping SM contributions to $C_{7 \gamma}$ and $C_{8 g}$.
Because the SM prediction of $B (B \rightarrow X_s \gamma)$ is
already consistent with the data,
there is less room for new physics in this case than in the $RL$
dominance scenario.
Nevertheless, we can get sizable change in $\SphiK$, satisfying
$B (B \rightarrow X_s \gamma)$.
Plots for this case can be found in Ref.~\cite{Kane:2002sp}.

\begin{figure*}[htbp]
  \centering
  \subfigure[$LL$ insertion case]
  {\includegraphics[height=5cm]{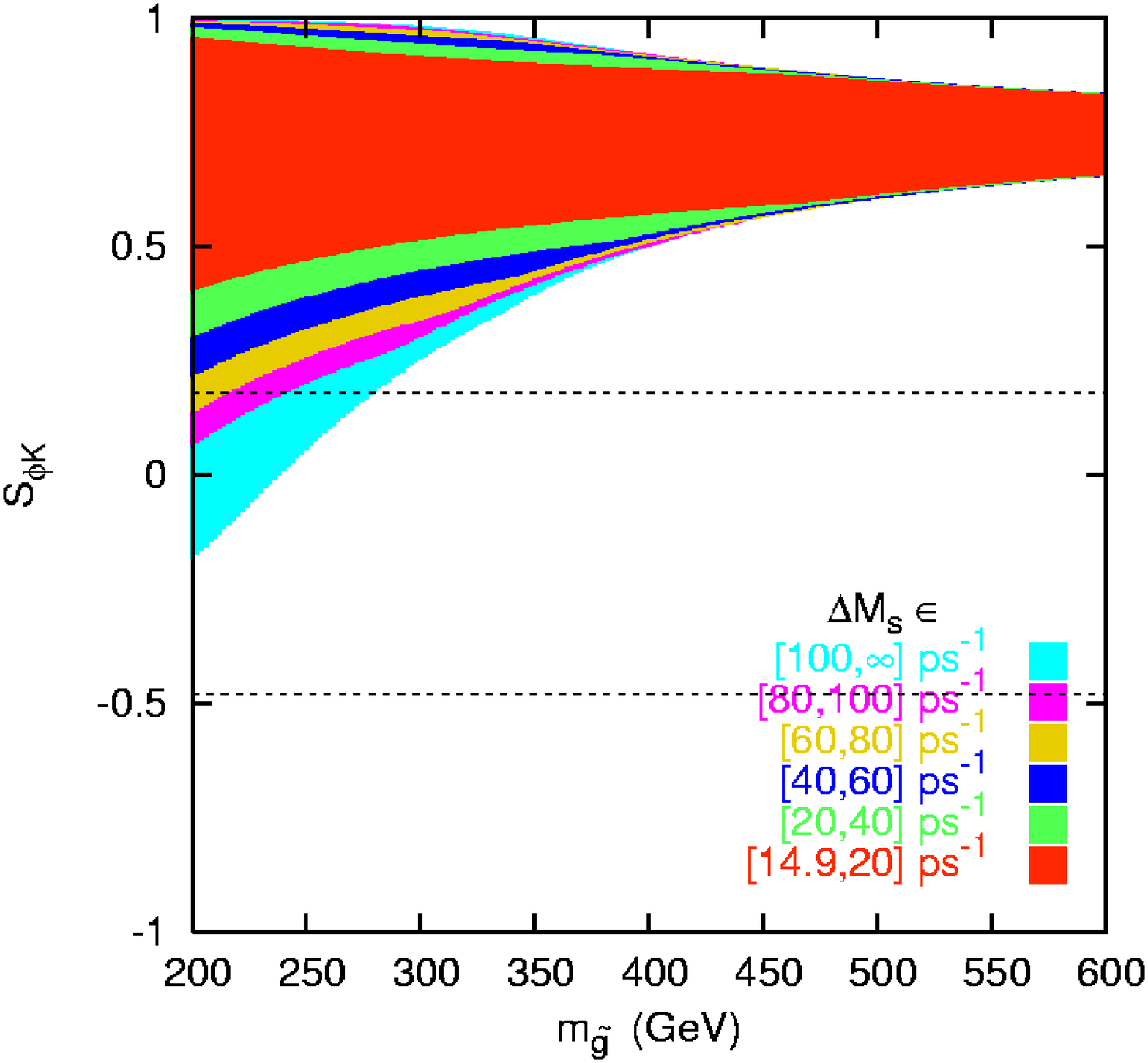}}
  \subfigure[$RR$ insertion case]
  {\includegraphics[height=5cm]{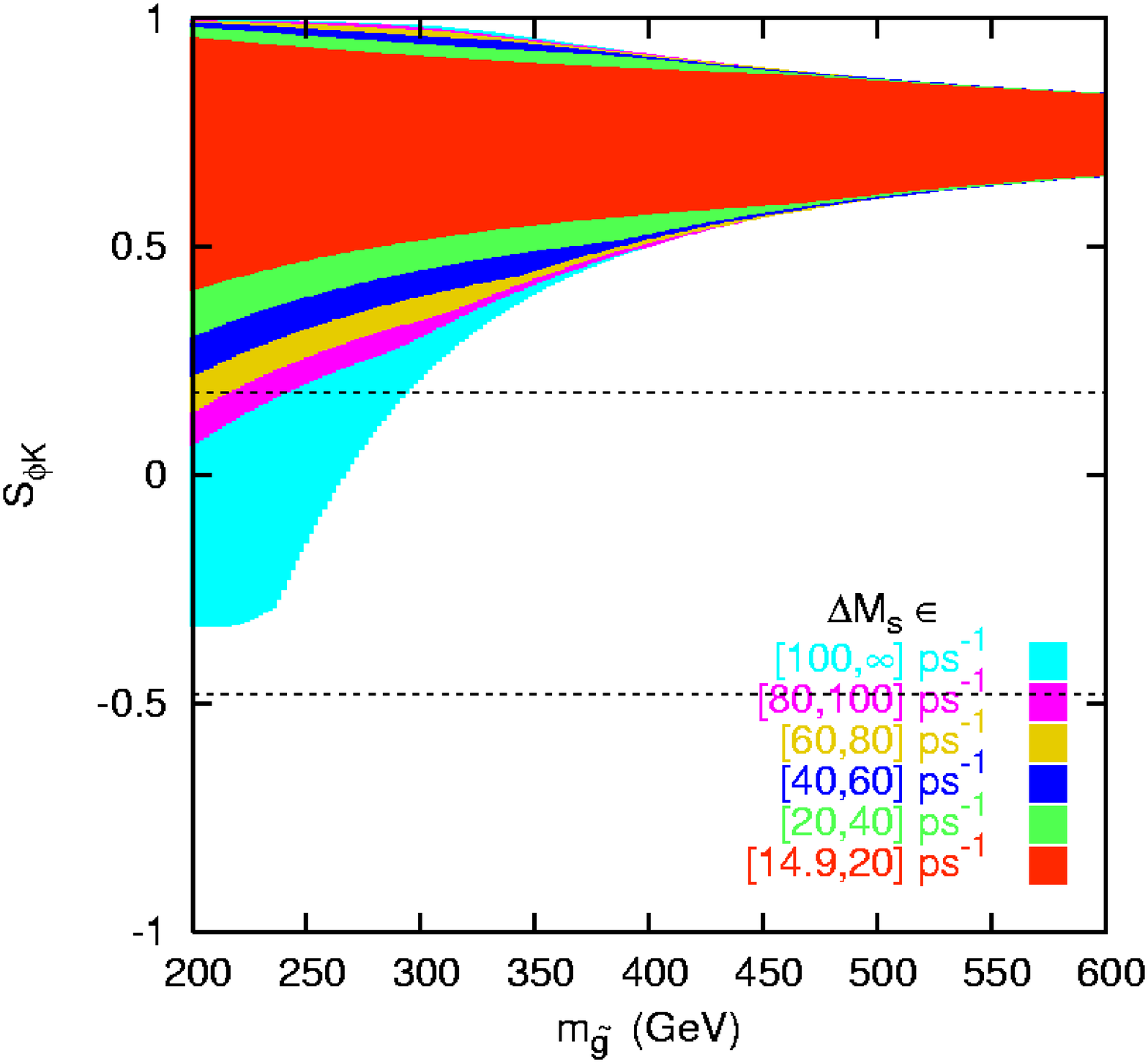}}
  \\
  \subfigure[$LR$ insertion case]
  {\includegraphics[height=5cm]{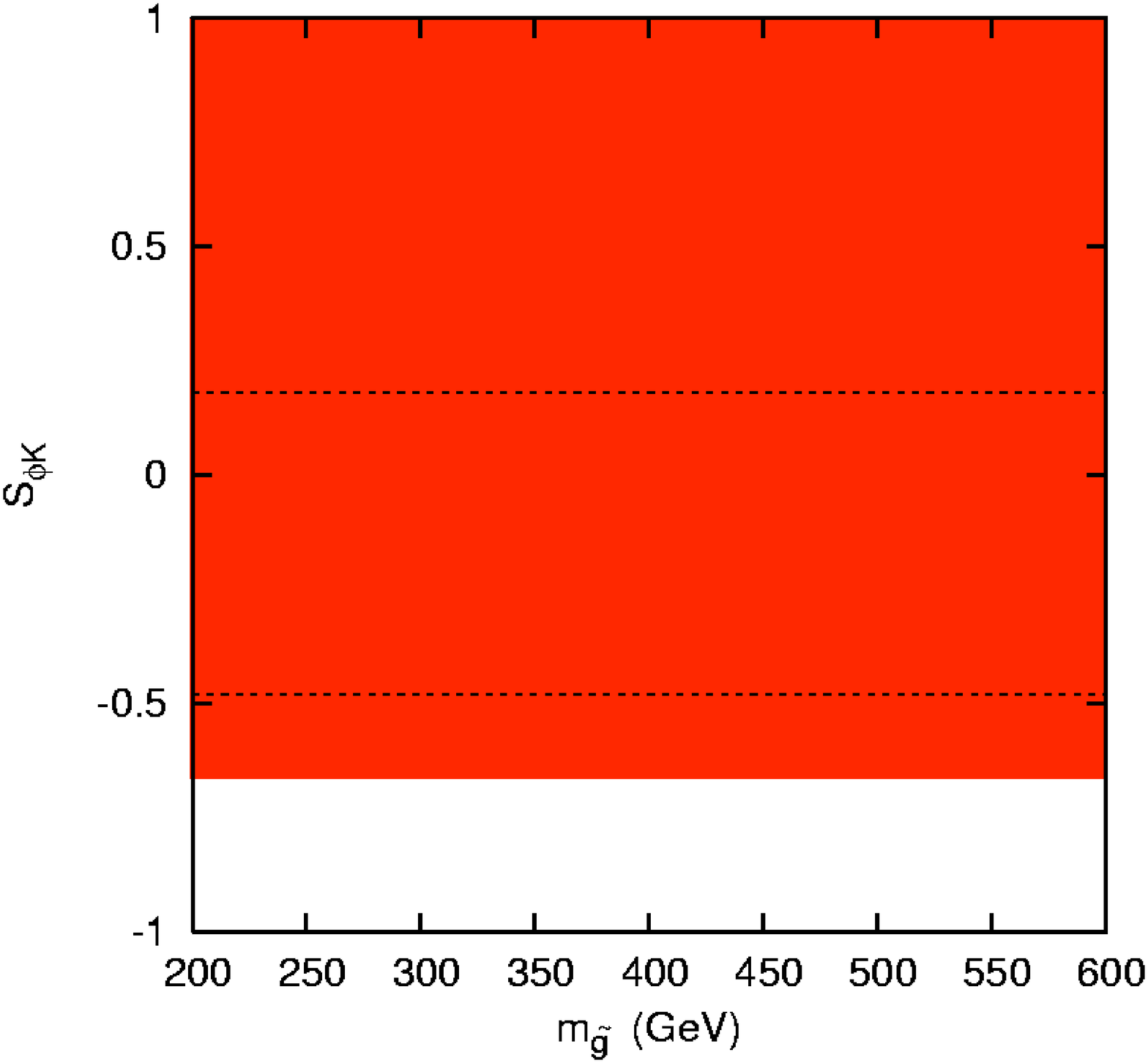}}
  \subfigure[$RL$ dominance case]
  {\includegraphics[height=5cm]{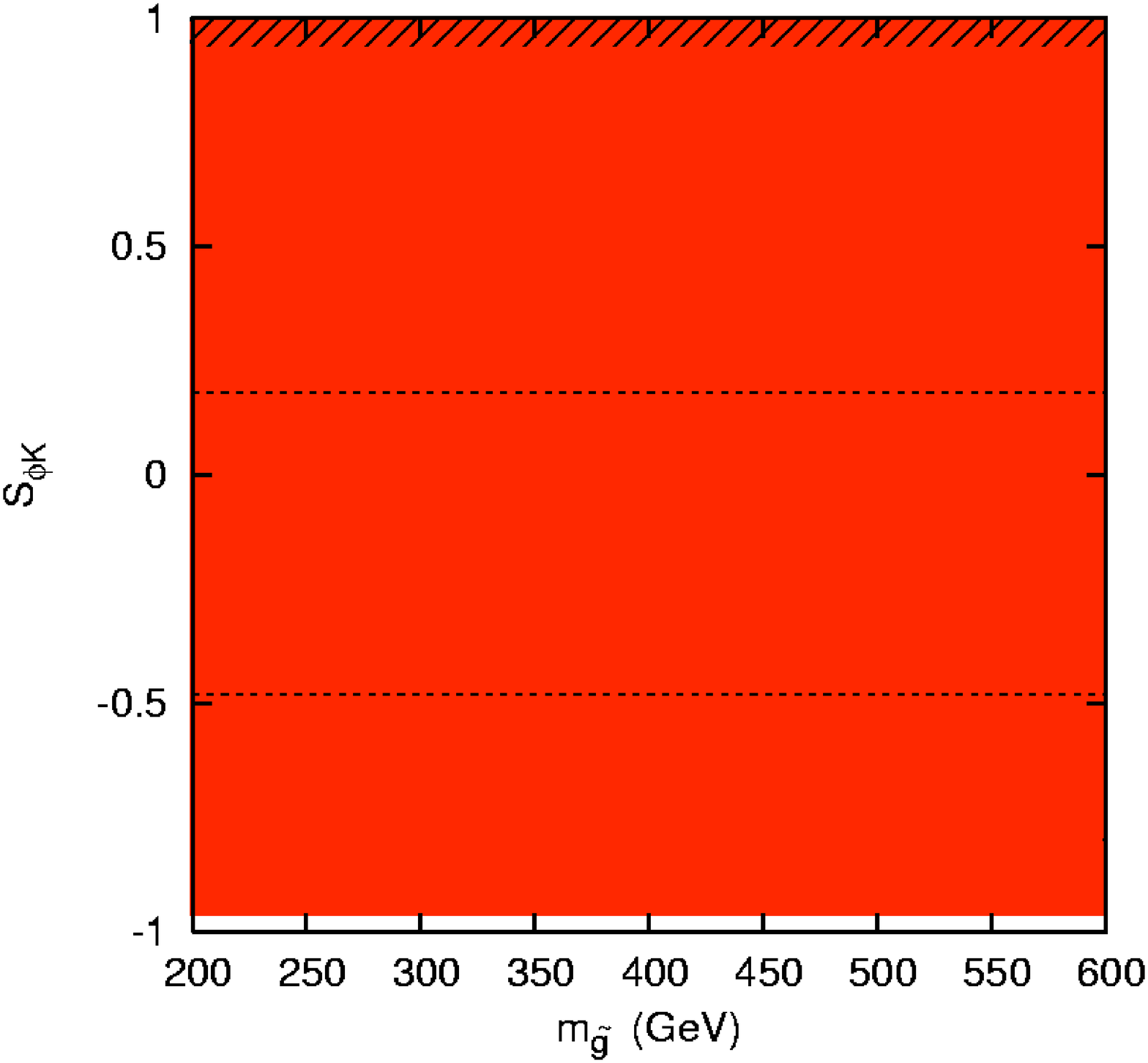}}
  \caption{Possible range of $\SphiK$ as a function of the gluino mass.
%    The ratio $x$ is always fixed to be 1.
    Different shades are used for different values of $\Delta M_s$.
    The dotted lines represent the current bounds on 
    $\SphiK$ at 1-$\sigma$ level.
  }
  \label{fig:mdep}
\end{figure*}

These results of analysis depend on the sparticle mass scale.
For a fixed value of MI parameter, SUSY effects
get enhanced as the sparticle masses decrease.
In Figs.~\ref{fig:mdep}, we show possible values of $\SphiK$
as functions of the gluino mass, $m_{\tilde{g}}$.
In these plots, 
we fix the ratio $x \equiv m_{\tilde{g}}^2 / \tilde{m}^2$ to be 1,
and scan over one of the four MI parameters imposing
constraints from $B (B \rightarrow X_s \gamma)$ and
$\Delta M_s$.
Figs.~\ref{fig:mdep}~(a) and (b) are for 
the chirality-preserving insertions, $LL$ and $RR$, respectively.
Here we see
a clear tendency of SUSY effects decoupling as the gluino mass
increases.
Note that our analysis shown in 
Figs.~\ref{fig:LLDB2}--\ref{fig:RLdDB1} is for $m_{\tilde{g}} = 400$ GeV.
The way to getting
$\SphiK$ consistent with the data
is to imagine the gluino mass $\sim 250$ GeV, which is
close to the current experimental lower bound.
On the other hand,
a chirality-flipping insertion, $LR$ or $RL$, shown
in Figs.~\ref{fig:mdep}~(c) and (d),
results in a range of $\SphiK$ that is independent of $m_{\tilde{g}}$.
This is explained in the following way in the $LR$ insertion case for example.
The dependence of $B \rightarrow X_s \gamma$ on the SUSY parameters
is only through the ratio of $( \delta_{LR}^d )_{23} / m_{\tilde{g}}$,
which precisely is the parameter that $\SphiK$ also depends on.
The set of $( \delta_{LR}^d )_{23} / m_{\tilde{g}}$ allowed by
$B (B \rightarrow X_s \gamma)$ is independent of $m_{\tilde{g}}$,
and so is $\SphiK$.
The $RL$ and $RL$ dominance cases are explained likewise.
In fact, this type of constant dependence would be true for
$LL$ or $RR$ insertion case as well,
if we allowed arbitrarily large size of $( \delta_{LL}^d )_{23}$ or
$( \delta_{RR}^d )_{23}$ and if we discarded $\Delta M_s$ constraint.

\begin{figure*}[htbp]
  \centering
  \subfigure[$\SphiK$ without hadronic uncertainties]
  {\includegraphics[height=5cm]{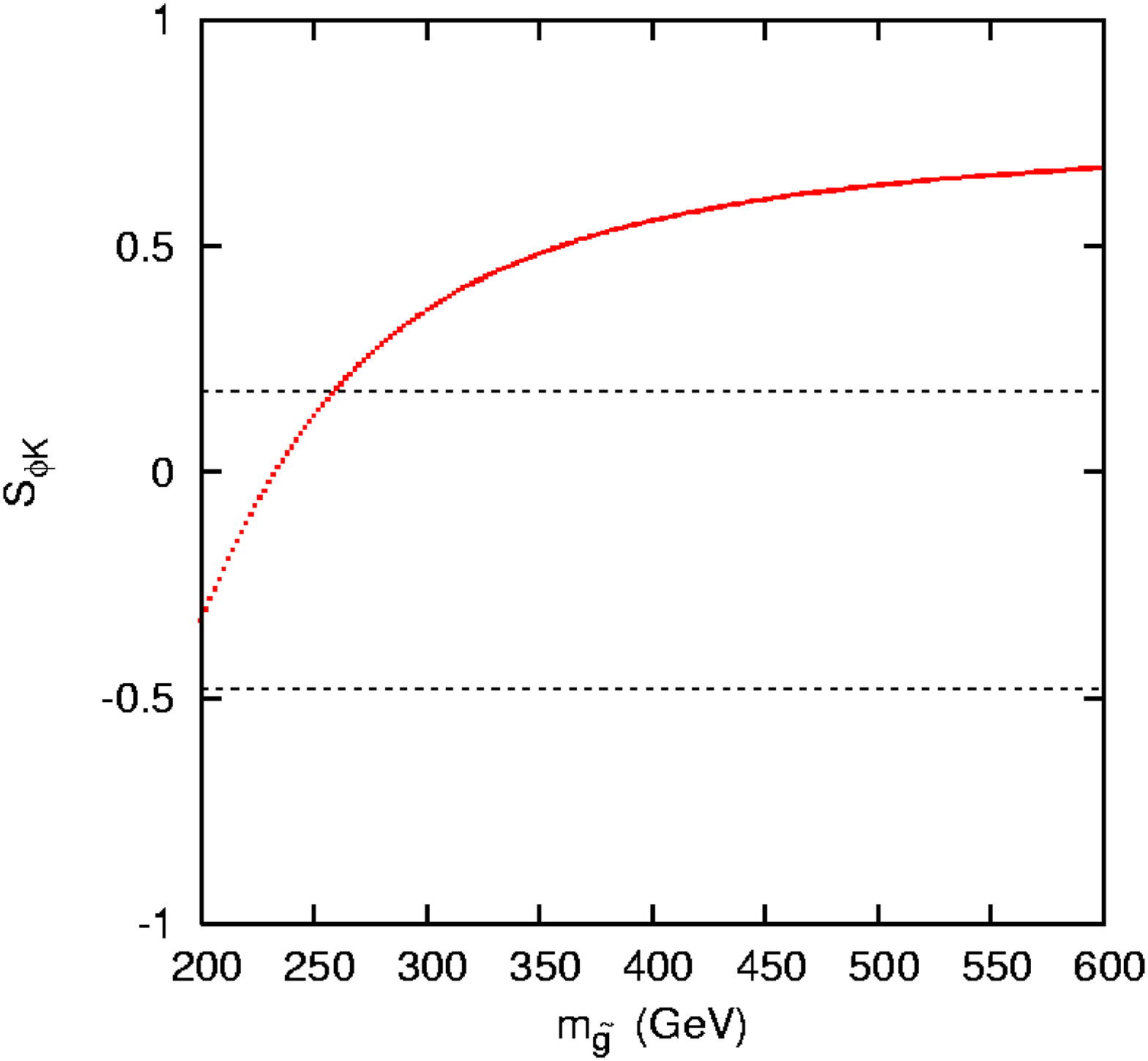}}
  \subfigure[$\SphiK$ with hadronic uncertainties]
  {\includegraphics[height=5cm]{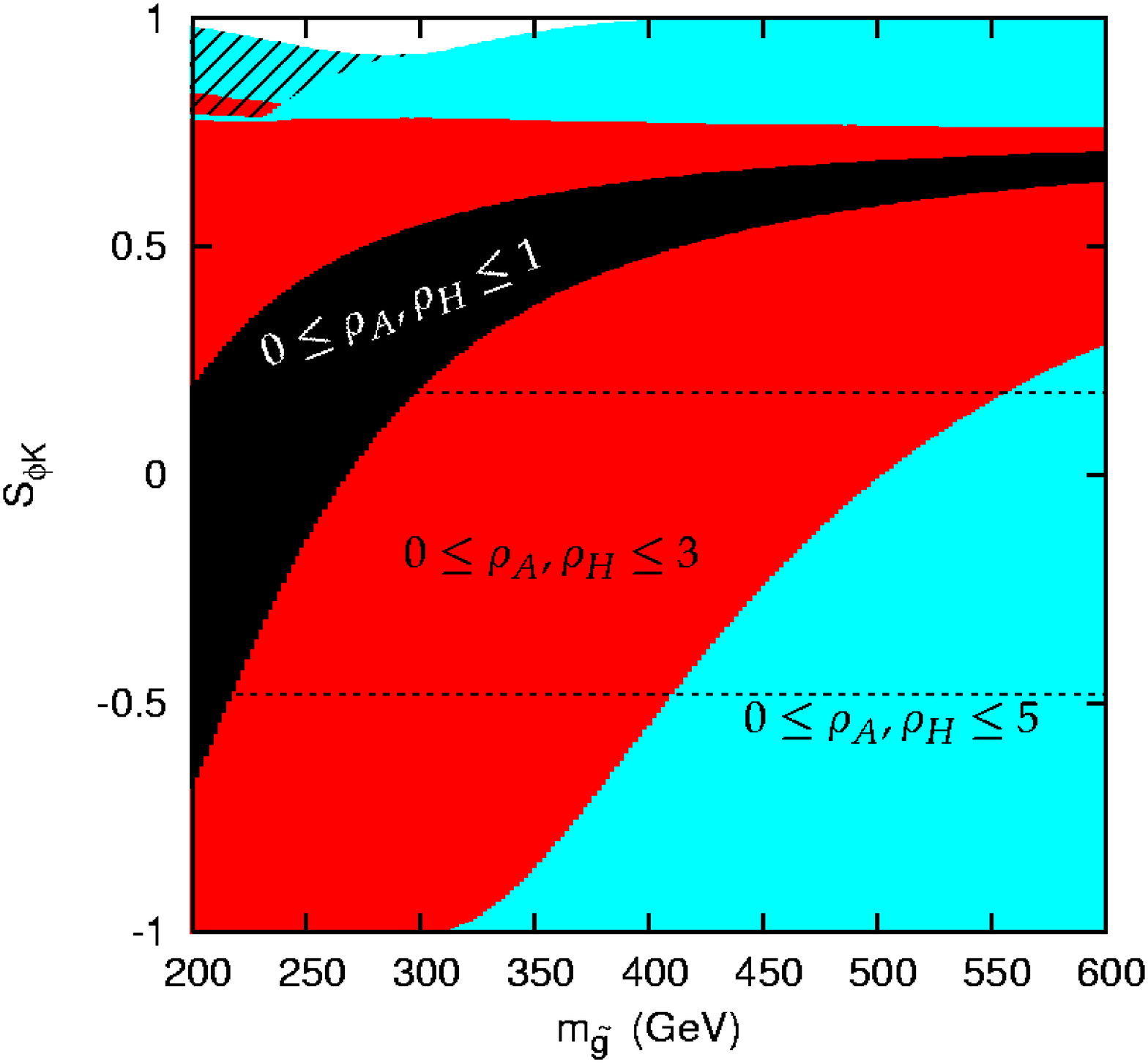}}
  \caption{Predictions of $\SphiK$ from 
    $(\delta^d_{23})_{RR} = 0.534 - 0.856 i $ as functions of 
    the gluino mass, (a) with $X_A = X_H = 0$, and (b)
    with varying $\varrho_A, \varrho_H$.
    The dotted lines represent the current bounds on 
    $\SphiK$ at 1-$\sigma$ level.
  In Fig.~(b), different shades are used for different ranges of
  $\varrho_A$ and $\varrho_H$.}
  \label{fig:uncer}
\end{figure*}

In the QCD factorization, 
divergences coming from endpoint of integral $\int dx/x$
are regulated in the form of
$X_{A(H)} = (1 + \varrho_{A(H)} e^{i \varphi_{A(H)}}) \ln (m_B / \Lambda_h)$,
where $\Lambda_h \sim 0.5$ GeV is the scale in hard-scattering.
Discussion so far has been carried out without considering 
any uncertainties coming from this regularization, but
we have to scan over the parameters $\varrho_{A,H}$ and $\varphi_{A,H}$
properly to take care of these
theoretical uncertainties.
In Fig.~\ref{fig:uncer}~(a), we show a curve 
of $\SphiK$ as a function of the gluino mass.
In this plot, we fix $x = 1$, and turn on only the $RR$ insertion 
with the value $(\delta^d_{23})_{RR} = 0.534 - 0.856 i$, which
minimizes $\SphiK$ at $m_{\tilde{g}} = 200$ GeV.
This curve is broadened into bands shown in Fig.~\ref{fig:uncer}~(b),
if we begin to scan over the aforementioned parameters.
In general, the uncertainty in the annihilation from $\varrho_A$
dominates over that in hard spectator scattering
from $\varrho_H$.
The black band is for the range $0 \le \varrho_A, \varrho_H \le 1$, 
and this is what the authors of Ref.~\cite{Beneke:1999br} recommend to use.
On this band, the tendency of decoupling SUSY effects with
increasing $m_{\tilde{g}}$ is still maintained.
However,
for the case of $0 \le \varrho_A, \varrho_H \le 5$,
which is displayed in light gray,
estimated theoretical uncertainties are so large that
almost every value of $\SphiK$ between $-1$ and 1 is predicted,
for any gluino mass between 200 GeV and 600 GeV.
Ref.~\cite{Ciuchini:2002uv} claims
that they get $\SphiK < 0$ from $LL$ or $RR$ insertion
with $m_{\tilde{g}} = 350$ GeV, disagreeing with us.
We suspect this is due to the difference in choosing intervals
for $\varrho_A$ and $\varrho_H$.
They use $0 \le \varrho_A \le 8$, and from Fig.~\ref{fig:uncer}~(b),
it is obvious that $\SphiK$ will have
every value between $-1$ and 1 for this interval.

% theoretical motivations
Let us think of theoretical motivations for
$(\delta_{LR}^d )_{23}$ or $(\delta_{LR}^d )_{23} \sim 10^{-2}$.
In many alignment models using flavor symmetry, there remain small
off-diagonal elements in the squark mass matrix in the super CKM basis.
They can cause $(\delta^d_{23})_{LL}$ or $(\delta^d_{23})_{RR}$
$\sim 10^{-2}$ with arbitrary complex phase.
This chirality preserving insertion can lead to an
induced $LR$ or $RL$ insertion,
\begin{equation*}
( \delta_{LR}^d )_{23}^{\rm ind} =  ( \delta_{LL}^d )_{23} \times 
{ m_b ( A_b - \mu \tan\beta ) \over \tilde{m}^2 }
\sim 10^{-2},
\end{equation*}
provided $\tan \beta$ is 
large enough so that $\mu \tan \beta \sim 30$ TeV.
This kind of double insertion mechanism has been used
in explaining both $\epsilon_K$ and $\epsilon' / \epsilon_K$ from
a single $CP$ violation source of $( \delta_{LL}^d )_{12}$ \cite{Baek:1999jq}.
We also constructed a model that
naturally gives the desired amount of $LR$ or $RL$ insertion
using intersecting D5 branes \cite{Kane:2002sp}.

% other decays
Since QCD penguin contributes to $b \rightarrow s \bar{q} q$
for $q = u, d, s, c, b$
at equal strength,
one may worry about what happens to other decays 
such as $B \rightarrow \eta' K_S,\ \pi K,\ K^+ K^- K_S$.
There are other 4-quark operators which contribute only to these decays
not affecting $\BtophiKs$,
% Therefore it may be the case that only $\BtophiKs$ is significantly 
% modified.
and changes in these decay modes are not definitely predictable
in the present study.
On the other hand,
there is a nice mechanism which enables us to control 
$B \rightarrow VP$ and $B \rightarrow PP$ modes independently.
Parity invariance tells us that
the transition amplitudes of these modes depend on the
new physics contribution in such a way that
$B \rightarrow VP\ (PP)$ mode depends on 
$\sum_i [C^\mathrm{NP}_i +(-)\ \widetilde{C}^\mathrm{NP}_i] \, \#_i$,
where $C^\mathrm{NP}_i$ and $\widetilde{C}^\mathrm{NP}_i$
are Wilson coefficients with opposite chiralities,
coming from new physics.
Hence
we can control $B \rightarrow \eta' K_S$ and $\BtophiKs$ separately,
and only the latter will change 
if we imagine a situation where
$(\delta_{LR}^d )_{23} = (\delta_{RL}^d )_{23}$, for instance.
If this is the case, predictions of other observables such as
$\CphiK$ and $\Acp$
will be different than the four cases we considered here.
After all, it requires further study 
to see changes in other decay modes in more general cases.

\begin{figure}[htbp]
  \centering
  \includegraphics{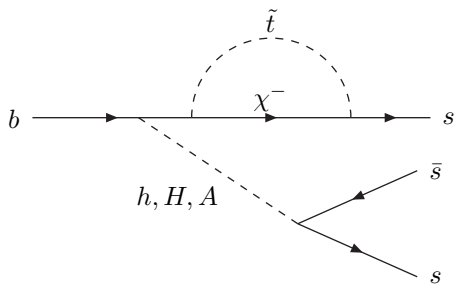}
  \caption{Higgs penguin diagram.}
  \label{fig:higgspen}
\end{figure}

% \begin{align}
%   \langle {VP} | H^\mathrm{NP}_\mathrm{eff} | B \rangle
%   =
%   \sum_i (C^\mathrm{NP}_i{+}\widetilde{C}^\mathrm{NP}_i) \, \#_i ,
%   \\
%   \langle {PP} | H^\mathrm{NP}_\mathrm{eff} | B \rangle
%   =
%   \sum_i (C^\mathrm{NP}_i{-}\widetilde{C}^\mathrm{NP}_i) \, \#_i ,
% \end{align}
% Higgs exchange
% From now, we switch to the minimal flavor violation scenario,
% which corresponds to completely vanishing squark MI parameters.
From now, we consider the neutral Higgs mediated contribution to
$\BtophiKs$.
% Analysis in the context of general MSSM can be found in \cite{Cheng:2003im}.
A typical diagram is shown in Fig.~\ref{fig:higgspen}.
This diagram
% The Higgs exchange diagram that contributes to $\BtophiKs$
contributes to $B_s \rightarrow \mu^+ \mu^-$ as well
as to $\BtophiKs$
if we replace the $s \bar{s}$ pair with $\mu^+ \mu^-$.
Once we impose the upper bound from CDF \cite{Abe:1998ah},
$B(B_s \rightarrow \mu^+ \mu^-) < 2.6 \times 10^{-6}$,
we find that $\SphiK > 0.71$.
This argument applies not only in minimal flavor violation scenario,
but also in general MSSM \cite{Kane:2002sp}.
As a result, Higgs exchange does not cause substantial change in $\SphiK$.

% conclusion
In conclusion,
we analyzed SUSY contributions to time dependent $CP$ asymmetry
in $\BtophiKs$.
As to the gluino-squark loop,
negative $\SphiK$ is more likely to come from $LR$ or $RL$ insertion
than from $LL$ or $RR$ insertion if the sparticle masses are not
close to the current experimental lower bound.
However, we may have chances to look for $LL$ or $RR$ insertion in
$\bsbsbar$ mixing.
Correlations among $\SphiK$, $\CphiK$, $\Acp$, $\Delta M_s$,
$\sin 2 \beta_s$ may help us discriminate among different
possibilities.
Constraint from nonleptonic decays such as $\BtophiKs$
is becoming as severe as that
from $B \rightarrow X_s \gamma$.
Higgs mediated FCNC cannot explain $\SphiK < 0$.
Finally, let us direct the reader to Ref.~\cite{Kane:2002sp} for
more complete discussion including
dilepton $CP$ asymmetry in $\Upsilon(4S)$ decay.

The author is grateful to G.~L.~Kane, P.~Ko, C.~Kolda, Haibin Wang,
and Lian-Tao Wang, for enjoyable collaboration.
This work was supported in part by KOSEF through CHEP at
Kyungpook National University and by the BK21 program.


\begin{thebibliography}{9}
\bibitem{browder}
T. Browder, talk at Lepton Photon 2003, Fermilab, Aug. 11-16, 2003.

%\cite{Abe:2003yt}
\bibitem{Abe:2003yt}
K.~Abe {\it et al.}  [Belle Collaboration],
%``Measurement of time-dependent CP-violating asymmetries in B0 $\to$ Phi K0(S), K+ K- K0(S), and eta' K0(S) decays,''
hep-ex/0308035.
%%CITATION = HEP-EX 0308035;%%

%\cite{Everett:2001yy}
\bibitem{Everett:2001yy}
L.~Everett, G.~L.~Kane, S.~Rigolin, L.~T.~Wang and T.~T.~Wang,
%``Alternative approach to b $\to$ s gamma in the uMSSM,''
JHEP {\bf 0201}, 022 (2002).
%[arXiv:hep-ph/0112126].
%%CITATION = HEP-PH 0112126;%%

%\cite{Stocchi:2000ps}
\bibitem{Stocchi:2000ps}
A.~Stocchi,
%``Review on B0 - anti-B0 mixing and B lifetimes measurements at CDF /  LEP / SLD,''
hep-ph/0010222.
%%CITATION = HEP-PH 0010222;%%

%\cite{Beneke:1999br}
\bibitem{Beneke:1999br}
M.~Beneke, G.~Buchalla, M.~Neubert and C.~T.~Sachrajda,
%``{QCD} factorization for B $\to$ pi pi decays: Strong phases and CP  violation in the heavy quark limit,''
Phys.\ Rev.\ Lett.\  {\bf 83}, 1914 (1999);
%[arXiv:hep-ph/9905312].
%%CITATION = HEP-PH 9905312;%%
%\cite{Beneke:2000ry}
%\bibitem{Beneke:2000ry}
%M.~Beneke, G.~Buchalla, M.~Neubert and C.~T.~Sachrajda,
%``QCD factorization for exclusive, non-leptonic B meson decays: General  arguments and the case of heavy-light final states,''
Nucl.\ Phys.\ B {\bf 591}, 313 (2000);
%[arXiv:hep-ph/0006124].
%%CITATION = HEP-PH 0006124;%%
%\cite{Beneke:2001ev}
%\bibitem{Beneke:2001ev}
%M.~Beneke, G.~Buchalla, M.~Neubert and C.~T.~Sachrajda,
%``QCD factorization in B $\to$ pi K, pi pi decays and extraction of  Wolfenstein parameters,''
Nucl.\ Phys.\ B {\bf 606}, 245 (2001).
%[arXiv:hep-ph/0104110].
%%CITATION = HEP-PH 0104110;%%

\bibitem{Kane:2002sp}
G.~L.~Kane, P.~Ko, H.~b.~Wang, C.~Kolda, J.-h.~Park and L.~T.~Wang,
%``B/d $\to$ Phi K(S) and supersymmetry,''
hep-ph/0212092;
%%CITATION = HEP-PH 0212092;%%
%\bibitem{Kane:2003zi}
%G.~L.~Kane, P.~Ko, H.~b.~Wang, C.~Kolda, J.~h.~Park and L.~T.~Wang,
%``B/d $\to$ Phi K(S) CP asymmetries as an important probe of supersymmetry,''
Phys.\ Rev.\ Lett.\  {\bf 90}, 141803 (2003).
%[arXiv:hep-ph/0304239].
%%CITATION = HEP-PH 0304239;%%

%\cite{Ciuchini:2002uv}
\bibitem{Ciuchini:2002uv}
M.~Ciuchini, E.~Franco, A.~Masiero and L.~Silvestrini,
%``b $\to$ s transitions: A new frontier for indirect SUSY searches,''
Phys.\ Rev.\ D {\bf 67}, 075016 (2003)
[Erratum-ibid.\ D {\bf 68}, 079901 (2003)].
%[arXiv:hep-ph/0212397].
%%CITATION = HEP-PH 0212397;%%

%\cite{Baek:1999jq}
\bibitem{Baek:1999jq}
S.~Baek, J.~H.~Jang, P.~Ko and J.-h.~Park,
%``Fully supersymmetric CP violations in the kaon system,''
Phys.\ Rev.\ D {\bf 62}, 117701 (2000);
%[arXiv:hep-ph/9907572];
%%CITATION = HEP-PH 9907572;%%
%\cite{Baek:2001kc}
%\bibitem{Baek:2001kc}
%S.~Baek, J.~H.~Jang, P.~Ko and J.~H.~Park,
%``Gluino-squark contributions to CP violations in the kaon system,''
Nucl.\ Phys.\ B {\bf 609}, 442 (2001).
%[arXiv:hep-ph/0105028].
%%CITATION = HEP-PH 0105028;%%

%\cite{Abe:1998ah}
\bibitem{Abe:1998ah}
F.~Abe {\it et al.}  [CDF Collaboration],
%``Search for the decays B/d0 $\to$ mu+ mu- and B/s0 $\to$ mu+ mu- in  p anti-p collisions at s**(1/2) = 1.8-TeV,''
Phys.\ Rev.\ D {\bf 57}, 3811 (1998).
%%CITATION = PHRVA,D57,3811;%%

\end{thebibliography}
\end{document}